# A Wearable RFID-Based Navigation System for the Visually Impaired

Fateme Zare
*Faculty of Electrical Engineering*
*K. N. Toosi University of Technology*
Tehran, Iran fatemezare79@email.kntu.ac.ir

Paniz Sedighi
*Faculty of Electrical and Computer Engineering*
*University of Alberta*
Alberta, Canada
Sedighi1@ualberta.ca

Mehdi Delrobaei
*Faculty of Electrical Engineering*
*K. N. Toosi University of Technology*
Tehran, Iran
delrobaei@kntu.ac.ir

*Abstract*—Recent studies have focused on developing advanced assistive devices to help blind or visually impaired people. Navigation is challenging for this community; however, devel- oping a simple yet reliable navigation system is still an unmet need. This study targets the navigation problem and proposes a wearable assistive system. We developed a smart glove and shoe set based on radio-frequency identification technology to assist visually impaired people with navigation and orientation in indoor environments. The system enables the user to find the directions through audio feedback. To evaluate the device's performance, we designed a simple experimental setup. The proposed system has a simple structure and can be personalized according to the user's requirements. The results identified that the platform is reliable, power efficient, and accurate enough for indoor navigation.

## I. INTRODUCTION

A recent global report by WHO indicates that at least 2.2 billion people struggle with vision impairment or blindness [1]. Visual impairment can be prevented in at least 50% of these cases. Also, the majority of vision impairment in low- and middle-income regions is four times higher than in high-income regions.

The visually impaired are often challenged with interacting with the surroundings, such as navigating unfamiliar environ- ments. As a result, they have lower workforce participation rates and may experience more inconvenience.

The white cane is commonly used by the blind and visually impaired. White canes are affordable and help the users to detect obstacles up to the knee level. However, they are not efficient in providing enough information about the environ- ment.

Assistive devices can enhance the quality of life of the visually impaired. Such devices may enable them to nav- igate independently, detect obstacles, and identify objects. Assistive technologies are also effective for rehabilitation, which improves the functioning of the visually impaired. Inventions such as electronic travel aids (ETAs) help them to improve their mobility. Although various assistive navigation devices are available, they are less popular among the visually impaired [2]. This fact indicates that further research should be conducted to enhance the capability and usability of assistive devices.

For this purpose, a practical design must be able to meet the following specifications: (1) simple and low-cost construction,
(2) lightweight, (3) reliable user interface, (4) power-efficient, and (5) acceptable accuracy.

In this work, we propose a system that includes a glove and a shoe based on radio-frequency identification (RFID) to meet all these requirements. The shoe and the glove provide non-contact data transfer between their transponders and the RFID passive tags, either on the floor or the surrounding objects. The user is then guided by a set of audio feedback generated based on a navigation algorithm and audio files saved on an onboard database.

This paper is organized as follows. Section II surveys related work, discussing the advantages and limitations of existing solutions. Section III describes the design and development of our proposed system. Section IV demonstrates the designed experiment and reports experimental results as well as the limitations of the system. Section V concludes the paper and proposes a future perspective.

## II. RELATED WORK

Current assistive devices employ different technologies to help the visually impaired. We categorized the solutions into four groups: RGB-D camera-based, ultrasonic and Infrared-based, GPS-based, and RFID-based methods.

### A. RGB-D camera-based methods

Barontini *et al.* [3] proposed a system based on wearable haptic technologies to help the visually impaired navigate and detect obstacles in indoor environments. The system consisted of an RGB-D camera, a microcontroller, and a wearable device. However, the device's usability seems questionable, and the camera may not be functional in dark environments.

Bai *et al.* [4] demonstrated a system including an RGB-D camera, a smartphone, IMU sensors, and earphones to receive the audio feedback. The device was designed for object recognition and localization in indoor and outdoor environments, but only indoor applications were achieved.

Aladre´n *et al.* [5] designed a simple RGB-D device for navigation and obstacle detection in indoor environments. They used sound map information and voice comments to guide the user.

Several researchers [6], [7] developed devices based on RGB-D cameras for object or face recognition. Although using machine vision algorithms, RGB-D cameras, stereoscopic, and binocular sensors provides a more accurate explanation of the environment, they need complex calculations and are expensive [8], [9].

*B. Ultrasound and Infrared-based methods*

Infrared (IR) based techniques are the most common positioning systems due to the availability of IR technology for various gadgets. An IR-based positioning system needs a line of sight link between the transmitter and receiver without any interference. The benefits of this technology are its small size and lightweight. The major limitations are the short range and maintenance cost [10].

NavGuide is an assistive device to aid visually impaired people in detecting obstacles. The device consisted of a low-power embedded system with ultrasonic sensors, vibration motors, and a power supply. The NavGuide provides audio feedback to the user using headphones [11].

The systems based on ultrasound technology are relatively low-cost, but the precision is lower than IR-based systems owing to the reflection influence. Additionally, this kind of system is always associated with other technologies, which may increase the cost of the system [12]. Furthermore, temperature, humidity, and high-frequency sounds can affect the measurements.

*C. GPS-based methods*

The GPS-based techniques are commonly used in outdoor navigation [13] since the GPS signal is weakened inside the buildings, and the accuracy is usually less than 15 meters [14]. Vela´zquez *et al.* [15] designed a system consisting of a smartphone, a microcontroller, and an RF module. The data is transmitted to the cloud server via an internet connection, and the user receives the foot tactile feedback to find the direction. Another research by Ramadhan [16] proposed a wearable device including ultrasonic sensors, a microcontroller, accelerometers, a GSM module, and a GPS.

*D. RFID-based methods*

The RFID sensing methods will become increasingly popular among researchers in different fields. The RFID sensor technology will be more utilized in biomedical areas [17]. Devipriya *et al.* [18] proposed an RFID-based smart store assistor for the visually impaired. This system consisted of three modules: product identifier, smart glove and smart trolley. In another work, Meshram *et al.* [19] provided NaveCane, an assistive device for autonomous orientation, to help the visually impaired.

Bruno Ando [20] presented a measurement strategy to assess an RFID-based navigation aid. The proposed system consisted of UHF RFID tags, an RFID reader, an antenna, and a microcontroller. They provided an index to evaluate the RFID transponder's sensitivity and measure its performance.

S. Alghamdi *et al.* [21] conducted two case studies to evaluate a combined technique of power attenuation and a received signal strength indicator using RFID. This technique was designed for both outdoor and indoor environments. However, their system had to be used with the combination of a white cane.

The poor user interface, functional complexity, inappropriate weight and size, and high price are reasons for the low popularity of the available assistive devices. Also, lack of a system that supports a 3D user interaction with the environment motivated us to address the limitations of existing navigating systems.

We developed a system consisting of an assistive glove and a shoe. The user navigates with the help of wireless communication between passive RFID tags and the transponders under the shoe and on the glove. The system has a user interface, including a keypad (input) and a set of headphones (outputs). The shoe mainly provides audio feedback to assist the visually impaired in navigating.

III. DESIGN AND DEVELOPMENT

*A. Design Concepts*

This section demonstrates the hardware components and the design and development procedure of the assistive navigation device. Our suggested platform is lightweight, easy to use, and low-cost. To develop a 3D interaction with the environment, we introduce a system consisting of two devices: an assistive glove for object recognition and a shoe for autonomous navigation.

After defining the design concept, we developed the assistive glove for object recognition. The glove's primary function is to identify passive RFID-tagged objects, relate audio recording messages to their unique IDs, and help the user navigate.

After testing the glove in different experiments and scenarios [22], we decided to design a system for helping the visually impaired in indoor navigation. Similarly, for the shoe, we defined a database of a room on the Raspberry Pi. Then we covered the room with RFID tags and embedded an RFID module under the shoe. The audio instructions were then generated after the user walked on the RFID passive tags to aid the user in navigating.

The shoe and the glove use independent onboard databases that can be utilized either separately or in an integrated mode. Furthermore, the shoe and the glove can communicate wirelessly through Raspberry Wi-Fi or Bluetooth to a PC or a smartphone. This configuration allows the user to create onboard databases where the required information about the target is saved.

According to Fig. 1, the system comprises two central parts: (1) the RFID tags located in specific locations or on specific objects, and (2) the wearable device worn by the user.

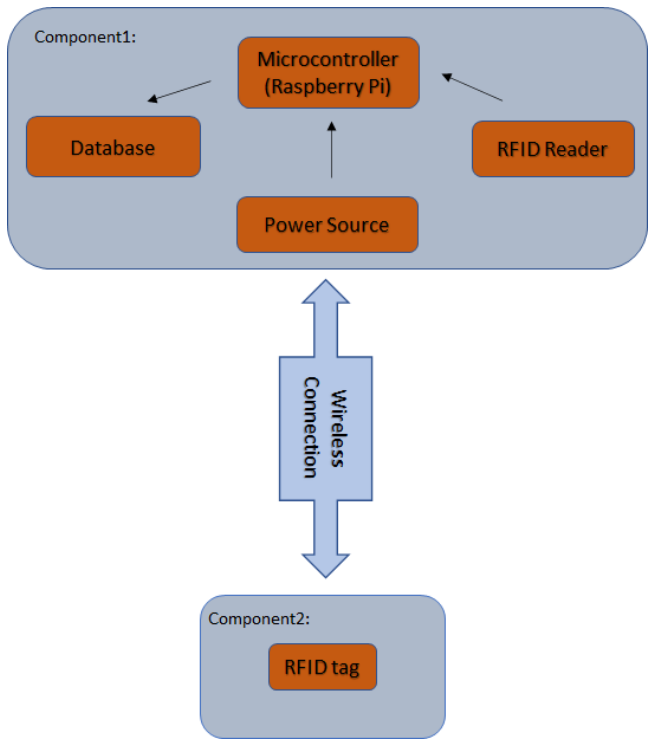

Fig. 1. The architecture of the system. Devices consist of RFID transponders, the processing units, the voice recorder (and player), the user interface, a power supply (up), and the RFID passive tags labeled in specific places or on objects (down).

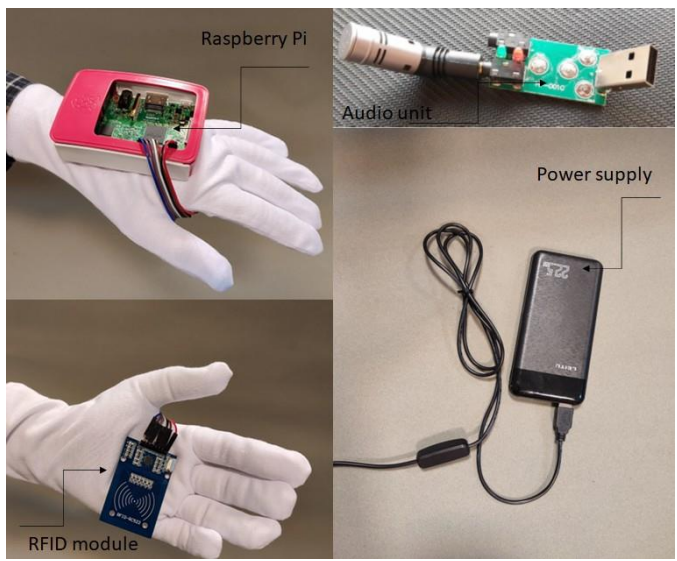

Fig. 2. The final prototype of the glove.

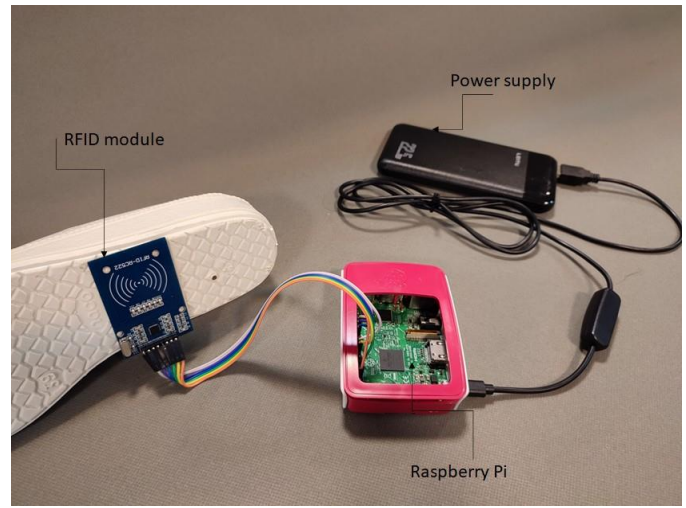

Fig. 3. The proposed prototype of the shoe.

### B. Prototype Development

The proposed platforms consist of five main components: the RFID transponder, the microcontroller, the voice player, the user interface, and the power supply (Figs. 2 and Fig. 3).This section provides the details of the navigation system's prototype.

*1) User interaction:* Based on the number of available tags in the room, the keypad takes multiple digits as input, where each number represents a unique tag on the floor. By entering a number and then pressing the hash key, the destination is set to the corresponding tag.

The star key can be used to restart the program. Aside from giving the user voice commands towards the destination, we designated a separate button to play audio recordings that describe the user's position. If users require additional information about their surroundings or nearby objects, they must press the "A" key. This kind of audio information exists only for certain tags that represent landmarks.

*2) The path-finding algorithm:* A graph can be formed, considering the arranged tags on the floor as nodes and the distances between them as edges. This graph displays a basic map of the environment. The nodes can also serve as landmarks, presenting a piece of equipment or a point of interest. The goal for the user could be to (1) find the shortest path to their desired destination, (2) avoid fixed obstacles, and (3) obtain general information about their position and surroundings. In order to achieve these goals, an onboard database and the shortest path algorithm are required.

*3) The shortest path algorithm:* In our design, each node is connected to a maximum of 6 neighboring nodes in a hexagonal pattern. All tags are equidistant, hence the weights of the edges are considered the same (equal to 1). This condition is true unless no edge can be identified due to the presence of a barrier or an obstacle, or it is comparatively hard to cross that edge. In such special cases, the weight is considered 2.

The complete graph includes all the vertices, their given names, their neighboring vertices, and the weights of the edges. An optimization algorithm is needed to approach the shortest path routing from any starting point to the desired target among the vertices in this graph. Fig. 4 explains the employed algorithm (based on Dijkstra's algorithm).

It is noted that we cannot give proper instructions if we do not know the user's orientation. The algorithm considers two consecutive tags to identify the orientation. In addition, there is no need to check whether the user has followed the instructions correctly, since Dijkstra's algorithm calculates a new path regardless of the user's previous step.

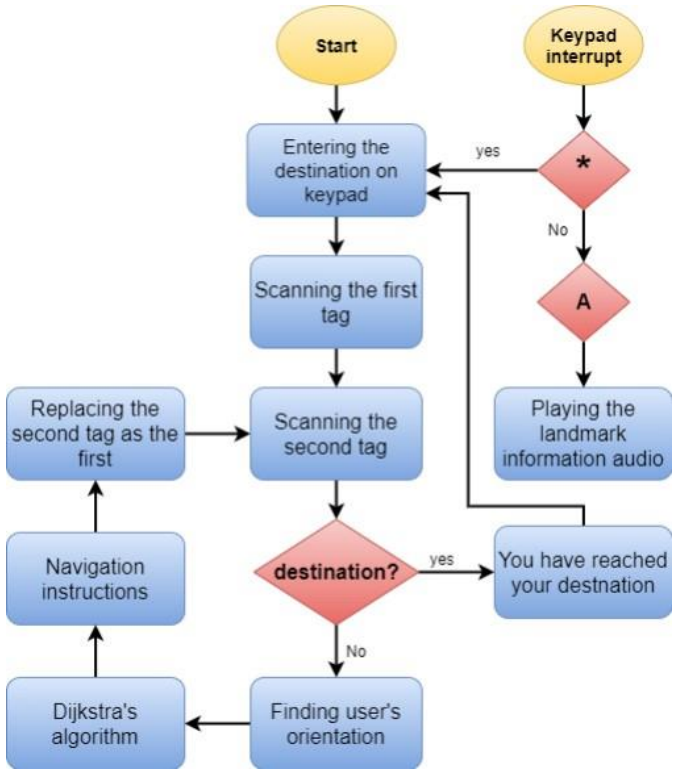

Fig. 4. The algorithm starts by entering the target tag number. After scanning two consecutive tags upon starting the program, the Dijkstra calculates the user's orientation. Then, the tags are scanned one by one as the user is given audio instructions. An interrupt button can restart the program, and another describes landmarks to the user.

*4) Database:* Storing and accessing such a comprehensive graph requires a database available at all times without any wireless connection. The database includes all the audio instructions and the characteristics of the room's plan, such as the arrangement of the tags, their position concerning each other, and the distances between them. The database was implemented using SQLite library in python. The SQLite library offers practical features such as expanding the database.

*5) The RFID Transponder:* The RFID module is the critical component. We were required to use an easy-to-use, affordable module. In this study, we used passive RFID tags with ISO 14443A standard and 13.56 MHz frequency and a 13.56 MHz MF-RC522 module. This transponder supports serial peripheral interface (SPI) protocol, I2C, and UART protocols and provides reliable communication by two-way data transfer at 424 Kbit/s. The transponder is 60 × 30 $mm^2$.

*6) The Processor:* To design the shoe, we implement the system on Arduino Nano 3, Raspberry Pi 3B+, and 4B. The Arduino platform is powered by an Atmega328 processor, which works at 16 MHz and 32 KB Flash Memory. We first implemented the system on the Arduino Nano3 to determine the defects. We needed a higher speed for processing the large tags up to 1000 numbers and appropriate storage for our audio files and databases.

Compared with Arduino Nano, Raspberry Pi 3B+ and 4B have faster 1.4 GHz and 1.5 GHz 64-bit quad-core processors. They also rely on removable micro SD cards, so they are suitable choices for our system. Since Raspberry Pi 4B and 3B+ approximately had the same power consumption, we selected Raspberry Pi 4B.

*7) The Audio Unit:* The user interacts with the system through audio feedback. The user is guided to navigate by playing audio messages while wearing the shoe and scanning RFID tags placed on the floor.

We first selected the WTV020-SD-16P sound module for the audio unit. After analyzing the system's performance, we added a USB sound card adapter and a microphone to the Raspberry Pi. This configuration enabled the proposed glove to record or play the audio messages. For the smart shoe, the audio instructions are played and received by a hands-free connection to the 3.5 mm audio jack.

*8) The Power Supply:* This study uses a power bank with a capacity of 10000 mAh to power the system.

## IV. EVALUATION AND EXPERIMENTAL RESULTS

A simple task was designed to evaluate the shoe's usability and performance. The details are as follows.

### A. Experimental Setup

A 4.1 × 3.2 $m^2$ room was considered to build the experimental setup. This room was supposed to represent a physician's office with two distinct areas: a reception area and a waiting room (including a coffee table, a water dispenser, and a vending machine). Twenty-four passive RFID tags with unique IDs covered the whole area. Fig. 6 illustrates the experimental setup. The tag placement enabled the user to avoid obstacles.

The tag placement was intended to help the users avoid obstacles. Therefore, the tags were placed in the vicinity of the obstacles, such as the coffee table. The location of the obstacles was initially registered on the onboard database (hence, no extra sensors were required). The system could provide two types of audio feedback. The first was the navigation instruc- tions, and the second one was the landmark announcement. Table I and Table II show the audio instructions. The audio cues were chosen to be brief and easy to follow. Fig. 5 shows the evaluation of the device in the real environment.

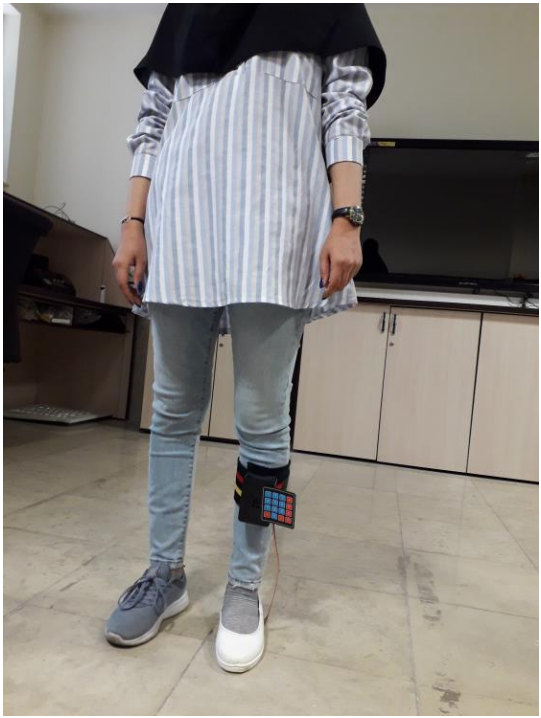

Fig. 5. The evaluation of the device.

TABLE I

ADUIO FEEDBACK FOR LANDMARKS

| Audio Feedback | Node |
| --- | --- |
| Office number 1 is located here. | A |
| This is the women's bathroom. | C |
| This is the men's bathroom. | D |
| This is the doctor's office. | I |
| This is office number 2. | J |
| There is a coffee table around. | M |
| This is the waiting area. | N |
| This is the reception table. | Q |
| There is a round table here. | R |
| This is the vending machine. | V |
| The receptionist is here. | X |

TABLE II
NAVIGATION ADUIO INSTRUCTIONS

| Audio Instruction | Direction |
| --- | --- |
| Walk straight ahead. | North |
| Turn to your 2 o'clock and keep walking slowly. | North-East |
| Turn to your 10 o'clock and keep walking slowly. | North-West |
| Make U-turn | South |
| Turn to your 4 o'clock and keep walking slowly. | South-East |
| Turn to your 8 o'clock and keep walking slowly. | South-West |

## B. Experimental Results

According to the experiment results, the RFID module's configuraton offers a detection range of nearly 5 cm with an effective angle of 60°. The maximum gain for the antenna is 5.5 dBi at a perpendicular angle. With multiple tags in the detection range, the transponder scans the most direct one. Also, nearly 30 tags can fit on a 10 $m^2$ surface with our proposed configuration. The tags are the same size as a credit card and can be covered by plaster or rug. The coverage does not significantly affect the reader's detection range. After the user scans a tag, it roughly takes up to 1.5 seconds to receive the audio instructions. For instance, the average time traveling from point "A" to point "Q" was 82 seconds.

The Raspberry Pi 4B has a peripheral current draw of 575mA (2.85 W) at idle and 600mA (3 W) while using one core. Meanwhile, the energy consumed by an RFID transponder is over 32 mA while reading passive tags. During the active state, when the device is scanning tags and playing audio recordings, the current draw adds up to over 650 mA (3.25 W), while at idle, it stays at a minimum of 575 mA. Given the 60% idle time average, the overall daily energy consumption is about 72.3 Wh. The device could be functional for 8 hours with a 10000 mAh power supply. Although energy efficiency was not our primary focus, the system seems power-efficient for intended applications. Table III summarizes the experimental results.

## C. Limitations

The relatively large distance between the tags was a limitation in this work. However, increasing the number of tags leads to more nodes, higher computational complexities, higher feedback delays, and slower walking speed. Such factors may cause difficulties for the user. Also, this work did not test the construction of a 3D tag placement and involvement of the glove in navigation (3D interaction with the environment).

## V. CONCLUSION AND FUTURE WORK

We presented the design details of a wearable device to assist people with visual impairment in indoor navigating in structured environments. The device showed efficiency in the designed experiment. The characteristics of the proposed system are listed as follows: 1) This system is self-supporting and is not dependent on any devices such as smartphones or the Internet, 2) due to its simple structure, the platform is relatively affordable for the user and, thus, can be available to the visually impaired community, 3) this device can be personalized according to the user's requirements.

The system's reliability and accuracy will be evaluated as part of our future work in different scenarios with multiple visually impaired users. We also plan to improve the user interface by adding a Braille keypad to the system. A smaller Raspberry version, Raspberry Zero, and flexible boards could also be employed. We also noted that the system must be water-resistant in some applications.

TABLE III
SUMARRY OF EXPERIMENTAL RESULTS

| Name | Result |
|---|---|
| RFID maximum detection Range and gain | 5 cm - 5.5 dBi |
| Average delay time for receiving instructions | 1.5 seconds |
| Average travelling velocity from "A" to "Q" | 5.46 cm/s |
| Overall daily energy consumption | 72.3 Wh |

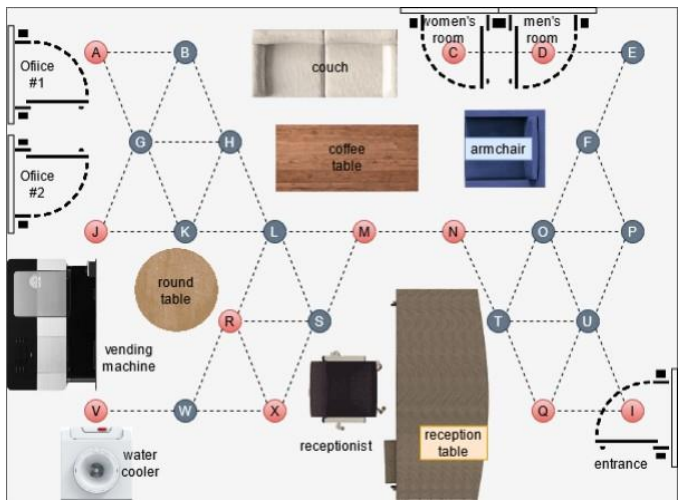

Fig. 6. The test room. The markers (landmarks) are shown in this room with red tags. The dashed line represents the possibility of passing that path. All tags have the same distance with one neighboring tag.